%
%
%
%
%
%
%
%

     
     
\font\twelverm=cmr10 scaled 1200    \font\twelvei=cmmi10 scaled 1200
\font\twelvesy=cmsy10 scaled 1200   \font\twelveex=cmex10 scaled 1200
\font\twelvebf=cmbx10 scaled 1200   \font\twelvesl=cmsl10 scaled 1200
\font\twelvett=cmtt10 scaled 1200   \font\twelveit=cmti10 scaled 1200
     
\skewchar\twelvei='177   \skewchar\twelvesy='60
     
     
\def\twelvepoint{\normalbaselineskip=12.4pt
  \abovedisplayskip 12.4pt plus 3pt minus 9pt
  \belowdisplayskip 12.4pt plus 3pt minus 9pt
  \abovedisplayshortskip 0pt plus 3pt
  \belowdisplayshortskip 7.2pt plus 3pt minus 4pt
  \smallskipamount=3.6pt plus1.2pt minus1.2pt
  \medskipamount=7.2pt plus2.4pt minus2.4pt
  \bigskipamount=14.4pt plus4.8pt minus4.8pt
  \def\rm{\fam0\twelverm}          \def\it{\fam\itfam\twelveit}%
  \def\sl{\fam\slfam\twelvesl}     \def\bf{\fam\bffam\twelvebf}%
  \def\mit{\fam 1}                 \def\cal{\fam 2}%
  \def\tt{\twelvett}
  \textfont0=\twelverm   \scriptfont0=\tenrm   \scriptscriptfont0=\sevenrm
  \textfont1=\twelvei    \scriptfont1=\teni    \scriptscriptfont1=\seveni
  \textfont2=\twelvesy   \scriptfont2=\tensy   \scriptscriptfont2=\sevensy
  \textfont3=\twelveex   \scriptfont3=\twelveex  \scriptscriptfont3=\twelveex
  \textfont\itfam=\twelveit
  \textfont\slfam=\twelvesl
  \textfont\bffam=\twelvebf \scriptfont\bffam=\tenbf
  \scriptscriptfont\bffam=\sevenbf
  \normalbaselines\rm}
     

     
\def\beginlinemode{\endmode
  \begingroup\parskip=0pt \obeylines\def\\{\par}\def\endmode{\par\endgroup}}
\def\beginparmode{\endmode
  \begingroup \def\endmode{\par\endgroup}}
\let\endmode=\par
{\obeylines\gdef\
{}}
\def\singlespace{\baselineskip=\normalbaselineskip}

\def\oneandahalfspace{\baselineskip=\normalbaselineskip
  \multiply\baselineskip by 3 \divide\baselineskip by 2}
\def\doublespace{\baselineskip=\normalbaselineskip \multiply\baselineskip by 2}

\newcount\firstpageno
\firstpageno=2
\footline={\ifnum\pageno<\firstpageno{\hfil}
                                          \else{\hfil\twelverm\folio\hfil}\fi}
\let\rawfootnote=\footnote              
\def\footnote#1#2{{\rm\singlespace\parindent=0pt\rawfootnote{#1}{#2}}}
\def\raggedcenter{\leftskip=4em plus 12em \rightskip=\leftskip
  \parindent=0pt \parfillskip=0pt \spaceskip=.3333em \xspaceskip=.5em
  \pretolerance=9999 \tolerance=9999
  \hyphenpenalty=9999 \exhyphenpenalty=9999 }
\def\dateline{\rightline{\ifcase\month\or
  January\or February\or March\or April\or May\or June\or
  July\or August\or September\or October\or November\or December\fi
  \space\number\year}}
\def\received{\vskip 3pt plus 0.2fill
 \centerline{\sl (Received\space\ifcase\month\or
  January\or February\or March\or April\or May\or June\or
  July\or August\or September\or October\or November\or December\fi
  \qquad, \number\year)}}
     
     
\hsize=6.5truein
\vsize=8.9truein
\parskip=\medskipamount
\twelvepoint            
\doublespace            
\overfullrule=0pt       
     
     
\def\preprintno#1{
 \rightline{\rm #1}}    
     
\def\title                      
  {\null\vskip 3pt plus 0.2fill
   \beginlinemode \doublespace \raggedcenter \bf}
     
\def\author                     
  {\vskip 3pt plus 0.2fill \beginlinemode
   \singlespace \raggedcenter}
     
\def\affil                      
  {\vskip 3pt plus 0.1fill \beginlinemode
   \oneandahalfspace \raggedcenter \sl}
     
\def\abstract                   
  {\vskip 3pt plus 0.3fill \beginparmode
   \doublespace \narrower ABSTRACT: }
     
\def\endtitlepage               
  {\endpage                     
   \body}
     
\def\body                       
  {\beginparmode}               

\def\subhead#1{                 
  \vskip 0.25truein             
  {\raggedcenter #1 \par}
   \nobreak\vskip 0.25truein\nobreak}
     
\def\refto#1{$|{#1}$}           
     
\def\references                 
  {\subhead{References}        
   \beginparmode
   \frenchspacing \parindent=0pt \leftskip=1truecm
   \parskip=8pt plus 3pt \everypar{\hangindent=\parindent}}
     
\gdef\refis#1{\indent\hbox to 0pt{\hss#1.~}}    
     
\gdef\journal#1, #2, #3, 1#4#5#6{             
    {\sl #1~}{\bf #2}, #3, (1#4#5#6)}           
     
\def\refstylenp{                
  \gdef\refto##1{ [##1]}                                
  \gdef\refis##1{\indent\hbox to 0pt{\hss##1)~}}        
  \gdef\journal##1, ##2, ##3, ##4 {                     
     {\sl ##1~}{\bf ##2~}(##3) ##4 }}
     
\def\refstyleprnp{              
  \gdef\refto##1{ [##1]}                                
  \gdef\refis##1{\indent\hbox to 0pt{\hss##1)~}}        
  \gdef\journal##1, ##2, ##3, 1##4##5##6{               
    {\sl ##1~}{\bf ##2~}(1##4##5##6) ##3}}

\def\np{\journal Nucl. Phys., }

\def\endreferences{\body}
     
\def\figurecaptions             
  { \beginparmode
   \subhead{Figure Captions}
}
     
\def\endfigurecaptions{\body}
     
\def\endpage                    
  {\vfill\eject}
     
\def\endpaper                   
  {\endmode\vfill\supereject}

\def\endit
  {\endpaper\end}

     
\def\ref#1{Ref. #1}                     
\def\Ref#1{Ref. #1}                     

\def\frac#1#2{{\textstyle{#1 \over #2}}}
\def\half{{\textstyle{ 1\over 2}}}

\def\sla{\raise.15ex\hbox{$/$}\kern-.57em}
\def\leaderfill{\leaders\hbox to 1em{\hss.\hss}\hfill}
\def\twiddle{\lower.9ex\rlap{$\kern-.1em\scriptstyle\sim$}}
\def\bigtwiddle{\lower1.ex\rlap{$\sim$}}
\def\gtwid{\mathrel{\raise.3ex\hbox{$>$\kern-.75em\lower1ex\hbox{$\sim$}}}}
\def\ltwid{\mathrel{\raise.3ex\hbox{$<$\kern-.75em\lower1ex\hbox{$\sim$}}}}
\def\square{\kern1pt\vbox{\hrule height 1.2pt\hbox{\vrule width 1.2pt\hskip 3pt
   \vbox{\vskip 6pt}\hskip 3pt\vrule width 0.6pt}\hrule height 0.6pt}\kern1pt}
\def\ucsb{Department of Physics\\University of California\\
Santa Barbara CA 93106-9530}

\def\m@th{\mathsurround=0pt }
\def\leftrightarrowfill{$\m@th \mathord\leftarrow \mkern-6mu
 \cleaders\hbox{$\mkern-2mu \mathord- \mkern-2mu$}\hfill
 \mkern-6mu \mathord\rightarrow$}
\def\overleftrightarrow#1{\vbox{\ialign{##\crcr
     \leftrightarrowfill\crcr\noalign{\kern-1pt\nointerlineskip}
     $\hfil\displaystyle{#1}\hfil$\crcr}}}


\font\titlefont=cmr10 scaled\magstep3

\def\martinstyletitle                      
  {\null\vskip 3pt plus 0.2fill
   \beginlinemode \doublespace \raggedcenter \titlefont}

\font\twelvesc=cmcsc10 scaled 1200  

\def\author                     
  {\vskip 3pt plus 0.2fill \beginlinemode
   \singlespace \raggedcenter\twelvesc}


\def\heading                            
  {\vskip 0.5truein plus 0.1truein      
   \beginparmode \def\\{\par} \parskip=0pt \singlespace \raggedcenter}

\def\subheading                         
  {\vskip 0.25truein plus 0.1truein     
   \beginlinemode \singlespace \parskip=0pt \def\\{\par}\raggedcenter}

\def\tag#1$${\eqno(#1)$$}

\def\align#1$${\eqalign{#1}$$}

\def\aligntag#1$${\gdef\tag##1\\{&(##1)\cr}\eqalignno{#1\\}$$
  \gdef\tag##1$${\eqno(##1)$$}}

\def\endaligntag{}

\def\overset #1\to#2{{\mathop{#2}\limits^{#1}}}
\def\underset#1\to#2{{\let\next=#1\mathpalette\undersetpalette#2}}
\def\undersetpalette#1#2{\vtop{\baselineskip0pt
\ialign{$\mathsurround=0pt #1\hfil##\hfil$\crcr#2\crcr\next\crcr}}}


\def\ref#1{Ref.~#1}                     
\def\Ref#1{Ref.~#1}                     
\def\[#1]{[\cite{#1}]}
\def\cite#1{{#1}}
\def\(#1){(\call{#1})}
\def\call#1{{#1}}
\def\taghead#1{}
\def\frac#1#2{{#1 \over #2}}
\def\half{{\frac 12}}

\def\12{{1\over2}}

\def\sla{\raise.15ex\hbox{$/$}\kern-.57em}
\def\leaderfill{\leaders\hbox to 1em{\hss.\hss}\hfill}
\def\twiddle{\lower.9ex\rlap{$\kern-.1em\scriptstyle\sim$}}
\def\bigtwiddle{\lower1.ex\rlap{$\sim$}}
\def\gtwid{\mathrel{\raise.3ex\hbox{$>$\kern-.75em\lower1ex\hbox{$\sim$}}}}
\def\ltwid{\mathrel{\raise.3ex\hbox{$<$\kern-.75em\lower1ex\hbox{$\sim$}}}}
\def\square{\kern1pt\vbox{\hrule height 1.2pt\hbox{\vrule width 1.2pt\hskip 3pt
   \vbox{\vskip 6pt}\hskip 3pt\vrule width 0.6pt}\hrule height 0.6pt}\kern1pt}
\def\tdot#1{\mathord{\mathop{#1}\limits^{\kern2pt\ldots}}}

\def\pmb#1{\setbox0=\hbox{#1}%
  \kern-.025em\copy0\kern-\wd0
  \kern  .05em\copy0\kern-\wd0
  \kern-.025em\raise.0433em\box0 }

\def \a{\rightarrow}
\def \d{\nabla}
\def \p{\partial}
\def \g{\gamma}
\def \t{\tilde}
\def \r{\hat r}

\catcode`@=11
\newcount\tagnumber\tagnumber=0

\immediate\newwrite\eqnfile
\newif\if@qnfile\@qnfilefalse
\def\write@qn#1{}
\def\writenew@qn#1{}
\def\w@rnwrite#1{\write@qn{#1}\message{#1}}
\def\@rrwrite#1{\write@qn{#1}\errmessage{#1}}

\def\taghead#1{\gdef\t@ghead{#1}\global\tagnumber=0}
\def\t@ghead{}

\expandafter\def\csname @qnnum-3\endcsname
  {{\t@ghead\advance\tagnumber by -3\relax\number\tagnumber}}
\expandafter\def\csname @qnnum-2\endcsname
  {{\t@ghead\advance\tagnumber by -2\relax\number\tagnumber}}
\expandafter\def\csname @qnnum-1\endcsname
  {{\t@ghead\advance\tagnumber by -1\relax\number\tagnumber}}
\expandafter\def\csname @qnnum0\endcsname
  {\t@ghead\number\tagnumber}
\expandafter\def\csname @qnnum+1\endcsname
  {{\t@ghead\advance\tagnumber by 1\relax\number\tagnumber}}
\expandafter\def\csname @qnnum+2\endcsname
  {{\t@ghead\advance\tagnumber by 2\relax\number\tagnumber}}
\expandafter\def\csname @qnnum+3\endcsname
  {{\t@ghead\advance\tagnumber by 3\relax\number\tagnumber}}

\def\equationfile{%
  \@qnfiletrue\immediate\openout\eqnfile=\jobname.eqn%
  \def\write@qn##1{\if@qnfile\immediate\write\eqnfile{##1}\fi}
  \def\writenew@qn##1{\if@qnfile\immediate\write\eqnfile
    {\noexpand\tag{##1} = (\t@ghead\number\tagnumber)}\fi}
}

\def\callall#1{\xdef#1##1{#1{\noexpand\call{##1}}}}
\def\call#1{\each@rg\callr@nge{#1}}

\def\each@rg#1#2{{\let\thecsname=#1\expandafter\first@rg#2,\end,}}
\def\first@rg#1,{\thecsname{#1}\apply@rg}
\def\apply@rg#1,{\ifx\end#1\let\next=\relax%
\else,\thecsname{#1}\let\next=\apply@rg\fi\next}

\def\callr@nge#1{\calldor@nge#1-\end-}
\def\callr@ngeat#1\end-{#1}
\def\calldor@nge#1-#2-{\ifx\end#2\@qneatspace#1 %
  \else\calll@@p{#1}{#2}\callr@ngeat\fi}
\def\calll@@p#1#2{\ifnum#1>#2{\@rrwrite{Equation range #1-#2\space is bad.}
\errhelp{If you call a series of equations by the notation M-N, then M and
N must be integers, and N must be greater than or equal to M.}}\else%
{\count0=#1\count1=#2\advance\count1 by1\relax\expandafter\@qncall\the\count0,%
  \loop\advance\count0 by1\relax%
    \ifnum\count0<\count1,\expandafter\@qncall\the\count0,%
  \repeat}\fi}

\def\@qneatspace#1#2 {\@qncall#1#2,}
\def\@qncall#1,{\ifunc@lled{#1}{\def\next{#1}\ifx\next\empty\else
  \w@rnwrite{Equation number \noexpand\(>>#1<<) has not been defined yet.}
  >>#1<<\fi}\else\csname @qnnum#1\endcsname\fi}

\let\eqnono=\eqno
\def\eqno(#1){\tag#1}
\def\tag#1$${\eqnono(\displayt@g#1 )$$}

\def\aligntag#1\endaligntag
  $${\gdef\tag##1\\{&(##1 )\cr}\eqalignno{#1\\}$$
  \gdef\tag##1$${\eqnono(\displayt@g##1 )$$}}

\def\eqalignno#1{\displ@y \tabskip\centering
  \halign to\displaywidth{\hfil$\displaystyle{##}$\tabskip\z@skip
    &$\displaystyle{{}##}$\hfil\tabskip\centering
    &\llap{$\displayt@gpar##$}\tabskip\z@skip\crcr
    #1\crcr}}

\def\displayt@gpar(#1){(\displayt@g#1 )}

\def\displayt@g#1 {\rm\ifunc@lled{#1}\global\advance\tagnumber by1
        {\def\next{#1}\ifx\next\empty\else\expandafter
        \xdef\csname @qnnum#1\endcsname{\t@ghead\number\tagnumber}\fi}%
  \writenew@qn{#1}\t@ghead\number\tagnumber\else
        {\edef\next{\t@ghead\number\tagnumber}%
        \expandafter\ifx\csname @qnnum#1\endcsname\next\else
        \w@rnwrite{Equation \noexpand\tag{#1} is a duplicate number.}\fi}%
  \csname @qnnum#1\endcsname\fi}

\def\ifunc@lled#1{\expandafter\ifx\csname @qnnum#1\endcsname\relax}

\let\@qnend=\end\gdef\end{\if@qnfile
\immediate\write16{Equation numbers written on []\jobname.EQN.}\fi\@qnend}

\catcode`@=12

\catcode`@=11
\newcount\r@fcount \r@fcount=0
\newcount\r@fcurr
\immediate\newwrite\reffile
\newif\ifr@ffile\r@ffilefalse
\def\w@rnwrite#1{\ifr@ffile\immediate\write\reffile{#1}\fi\message{#1}}

\def\writer@f#1>>{}
\def\referencefile{
  \r@ffiletrue\immediate\openout\reffile=\jobname.ref%
  \def\writer@f##1>>{\ifr@ffile\immediate\write\reffile%
    {\noexpand\refis{##1} = \csname r@fnum##1\endcsname = %
     \expandafter\expandafter\expandafter\strip@t\expandafter%
     \meaning\csname r@ftext\csname r@fnum##1\endcsname\endcsname}\fi}%
  \def\strip@t##1>>{}}

\def\citeall#1{\xdef#1##1{#1{\noexpand\cite{##1}}}}
\def\cite#1{\each@rg\citer@nge{#1}}

\def\each@rg#1#2{{\let\thecsname=#1\expandafter\first@rg#2,\end,}}
\def\first@rg#1,{\thecsname{#1}\apply@rg}	
\def\apply@rg#1,{\ifx\end#1\let\next=\relax
\else,\thecsname{#1}\let\next=\apply@rg\fi\next}

\def\citer@nge#1{\citedor@nge#1-\end-}	
\def\citer@ngeat#1\end-{#1}
\def\citedor@nge#1-#2-{\ifx\end#2\r@featspace#1 
  \else\citel@@p{#1}{#2}\citer@ngeat\fi}	
\def\citel@@p#1#2{\ifnum#1>#2{\errmessage{Reference range #1-#2\space is bad.}%
    \errhelp{If you cite a series of references by the notation M-N, then M and
    N must be integers, and N must be greater than or equal to M.}}\else%
{\count0=#1\count1=#2\advance\count1 by1\relax\expandafter\r@fcite\the\count0,%
  \loop\advance\count0 by1\relax
    \ifnum\count0<\count1,\expandafter\r@fcite\the\count0,%
  \repeat}\fi}

\def\r@featspace#1#2 {\r@fcite#1#2,}	
\def\r@fcite#1,{\ifuncit@d{#1}
    \newr@f{#1}%
    \expandafter\gdef\csname r@ftext\number\r@fcount\endcsname%
                     {\message{Reference #1 to be supplied.}%
                      \writer@f#1>>#1 to be supplied.\par}%
 \fi%
 \csname r@fnum#1\endcsname}
\def\ifuncit@d#1{\expandafter\ifx\csname r@fnum#1\endcsname\relax}%
\def\newr@f#1{\global\advance\r@fcount by1%
    \expandafter\xdef\csname r@fnum#1\endcsname{\number\r@fcount}}

\let\r@fis=\refis			
\def\refis#1#2#3\par{\ifuncit@d{#1}
   \newr@f{#1}%
   \w@rnwrite{Reference #1=\number\r@fcount\space is not cited up to now.}\fi%
  \expandafter\gdef\csname r@ftext\csname r@fnum#1\endcsname\endcsname%
  {\writer@f#1>>#2#3\par}}

\def\ignoreuncited{
   \def\refis##1##2##3\par{\ifuncit@d{##1}%
    \else\expandafter\gdef\csname r@ftext\csname r@fnum##1\endcsname\endcsname%
     {\writer@f##1>>##2##3\par}\fi}}

\def\r@ferr{\endreferences\errmessage{I was expecting to see
\noexpand\endreferences before now;  I have inserted it here.}}
\let\r@ferences=\references
\def\references{\r@ferences\def\endmode{\r@ferr\par\endgroup}}

\let\endr@ferences=\endreferences
\def\endreferences{\r@fcurr=0
  {\loop\ifnum\r@fcurr<\r@fcount
    \advance\r@fcurr by 1\relax\expandafter\r@fis\expandafter{\number\r@fcurr}%
    \csname r@ftext\number\r@fcurr\endcsname%
  \repeat}\gdef\r@ferr{}\endr@ferences}


\let\r@fend=\endpaper\gdef\endpaper{\ifr@ffile
\immediate\write16{Cross References written on []\jobname.REF.}\fi\r@fend}

\catcode`@=12

\citeall\refto		
\citeall\ref		%
\citeall\Ref		%

\def\Tr{\mathop{\rm Tr}\nolimits}
\def\gi{ g^{-1}}
\def\dz{\partial_{\scriptscriptstyle +}}
\def\db{\partial_{\scriptscriptstyle -}}
\def\Az{A_{\scriptscriptstyle +}}
\def\Ab{A_{\scriptscriptstyle -}}
\def\+{{\scriptscriptstyle +}}
\def\-{{\scriptscriptstyle -}}
\def\ints{\int_{\Sigma} d^2 \sigma \,}
\def\R{{\bf R}}
\def\slr{SL(2,\R)}
\def\jthree{\pmatrix{1&0\cr0&-1\cr}}
\def\RN{Reissner--Nordstr\"om}


\font\titlefont=cmr10 scaled \magstep3
\def\bigtitle{\null\vskip 3pt plus 0.2fill \beginlinemode \doublespace
\raggedcenter \titlefont}

\gdef\journal#1, #2, #3, 1#4#5#6{
{\sl #1~}{\bf #2}, #3 (1#4#5#6)}

\singlespace
\preprintno{UCSBTH-91-39}
\preprintno{July, 1991}
\doublespace
\bigtitle {Exact Black String Solutions in Three Dimensions}
\bigskip
\author James H. Horne and Gary T. Horowitz
\affil\ucsb
\centerline{jhh@cosmic.physics.ucsb.edu}
\centerline{gary@cosmic.physics.ucsb.edu}
\abstract
A family of exact conformal field theories is constructed which
describe charged black strings in three dimensions. Unlike previous
charged black hole or extended black hole solutions in string theory,
the low energy spacetime metric has a regular inner horizon (in addition to
the event horizon) and a
timelike singularity. As the charge to mass ratio approaches unity,
the event horizon remains but the singularity disappears.
						    
\endtitlepage

\baselineskip=16pt

\subhead{\bf 1. Introduction}

In a recent paper~[\cite{Horowitz}], it was shown that string theory
has a rich variety of solutions describing extended objects surrounded
by event horizons. In particular there are black string solutions in
ten dimensions characterized by three parameters: the mass and axion
charge per unit length, and the asymptotic value of the dilaton. These
solutions were obtained by solving the low energy string equations of
motion. Although this is sufficient to establish the existence of
exact solutions with these qualitative features, it was not clear how
to construct directly the conformal field theory with these
properties.

Witten has recently shown~[\cite{Witten}] that a simple gauged 
WZW model~[\cite{Gawedzki},\cite{Nemeschansky}] yields a two
dimensional black hole. This raises the possibility of
using similar constructions to find exact conformal field theories
corresponding to higher dimensional black holes or extended black
holes. (The conformal field theory associated with an extremal limit
of the charged black fivebranes has recently been
found~[\cite{Giddings}].) In this paper we will show that a simple
extension of Witten's construction yields  three dimensional charged
black strings. These solutions are also characterized by three parameters:
the mass $M$ and axion charge $Q$ per unit length, and a constant $k$
related to the asymptotic
value of the derivative of the dilaton. The low energy metric, antisymmetric
tensor, and dilaton take the form:
$$  ds^2 = - \left(1-{M\over r}\right) dt^2
           + \left(1-{Q^2\over M r}\right)dx^2
           + \left(1-{M\over r}\right)^{-1} \left(1-{Q^2\over Mr}\right)^{-1}
             {k \, dr^2 \over 8 r^2} $$
$$ \eqalign{ H_{rtx} & = Q/ r^2 \cr
          \Phi & = \ln r + {1 \over 2} \ln {k \over 2}. \cr} \eqno(final) $$

Perhaps the most important feature of these solutions is that their
global structure is qualitatively different from all previously
discussed string solutions. The existing solutions describing black
holes~[\cite{Witten}], charged black
holes~[\cite{Gibbons},\cite{Maeda},\cite{Ivanov},\cite
{Garfinkle},\cite{Ishibashi}],
and extended black holes~[\cite{Horowitz}] (away from their extremal
values) are all qualitatively similar to the Schwarzschild solution of
general relativity: The low energy spacetime metrics have an event
horizon and a spacelike singularity inside. In contrast, when $0 < |Q|
< M$, our black strings are similar to the \RN\ solution which
describes charged black holes in general relativity. In addition to
the event horizon, there is a second inner horizon and a timelike
singularity. When $|Q| = M$, we will show that the black string
solutions have the unusual property of possessing an event horizon,
but no singularity!  Finally, when $|Q| >M$, we will see that the
spacetime has neither a horizon nor a curvature singularity.  However
there is in general a conical singularity which can be removed by
identifying the $x$ coordinate with an appropriate period. This
changes the spacetime at infinity from $\R^3$ to $\R^2 \times S^1$. In
a certain limit, our solution reduces to the two dimensional {\it
Euclidean} black hole solution with a time direction added.

Although the three dimensional black strings are  most naturally described
in terms of the string metric (the metric appearing in the sigma model),
it is also of interest to consider the rescaled Einstein metric (with the
standard Einstein-Hilbert action)\footnote*{Since the two dimensional
Einstein action is a topological invariant, this is not possible for the
two dimensional black hole.}. We will see that the Einstein metric
also describes a black string in an asymptotically flat spacetime. But it
 is not static! There is still a timelike symmetry outside the event horizon,
 but it resembles a boost at infinity rather than a time translation.

\subhead{\bf 2. Derivation of Black String Solutions}

We now describe the conformal field theory construction which yields
the black strings.  Since our target space is going to have Lorentz
signature, we will use a Lorentz metric $ds^2 = 2d\sigma_\+
d\sigma_\-$ on the world sheet $\Sigma$. If $g$ is an element of a
group $G$, then the ungauged Wess--Zumino--Witten action can be
written
$$ L(g) = {k \over 4 \pi} \ints
                     \Tr ( \gi \dz g \gi \db g )
          - {k \over 12 \pi} \int_B \Tr ( \gi dg \wedge \gi dg \wedge \gi dg )
               \, ,  \eqno(Lg) $$
where $B$ is a three manifold with boundary $\Sigma$.

We are interested in gauging  a one dimensional subgroup $H$ of the  symmetry 
group  
of eq.~\(Lg), with action $g \rightarrow h g h$. We can make this
global symmetry
local by introducing a gauge field $A_i$ which takes values in the
Lie algebra of $H$. If $\epsilon$
is an infinitesimal gauge parameter, then the local axial
symmetry is generated by
$$ \delta g = \epsilon g + g \epsilon , \> \delta A_i = - \partial_i \epsilon
                                                 \, . \eqno(symm)$$
This local axial symmetry is a symmetry of the gauged WZW action
$$ L(g,A) = L(g) + {k \over 2 \pi} \ints
           \Tr ( \Az \db g \gi + \Ab \gi \dz g + \Az \Ab
                    + \Az g \Ab \gi ) \, . \eqno(LgA) $$

Witten showed that if $G$ is $\slr$ and $H$ is the subgroup
generated by $\jthree$, then the gauged WZW action
describes a black hole in two dimensions. We now generalize this construction
by adding one free boson $x$ to the action, which is equivalent to
letting $G = \slr \times \R$. We then gauge the one dimensional
subgroup generated
by $\jthree$ together with a translation of $x$.
To be explicit, parameterize the group manifold
of $\slr$ by
$$ g_{\slr} = \pmatrix{ a & u \cr -v & b \cr} \>\>{\rm with}\>\> a b + u v = 1
                                . \eqno(slrpar)$$
This gives the ungauged action
$$\eqalign{ L(g) = & -{k \over 4 \pi} \ints ( 
            \dz u \db v + \db u \dz v + \dz a \db b + \db a \dz b) \cr
          & + {k \over 2 \pi} \ints
                                     \log u (\dz a \db b - \db a \dz b)
   + {1 \over \pi} \ints \dz x \db x ,\cr} \eqno(ungauged) $$
We then gauge
$$\delta a = 2 \epsilon a , \>
  \delta b = - 2 \epsilon b, \>
  \delta u = \delta v = 0 , \>
  \delta x = 2 \epsilon c , \>
  \delta A_i = - \partial_i \epsilon .  \eqno(symexp)$$
where $c$ is an arbitrary constant\footnote*{A related construction with
a compactified $x$ has been used to obtain two dimensional charged black
holes~[\cite{Ishibashi}].}.
The full action is now
$$\eqalign{ L(g,A) = L(g) + {k \over 2 \pi} \ints 
   & \Az ( b \db a - a \db b - u \db v + v \db u + {4 c \over k} \db x ) \cr
 + & \Ab ( b \dz a - a \dz b + u \dz v - v \dz u + {4 c \over k} \dz x ) \cr
         + & 4 \Az \Ab (1 + {2 c^2 \over k} - u v) \>. \cr} \eqno(LgAtwo) $$
As in ref.~[\cite{Witten}],
we can now gauge fix by setting $a = \pm b$, depending on
the sign of $1 - u v$.
After making this gauge choice and eliminating $A$ the action becomes
$$\eqalign{ L = & - {k \over 8 \pi} \ints { \lambda v^2 \dz u \db u
                + \lambda u^2 \dz v \db v
          + (2 - 2 u v + 2 \lambda - \lambda u v)(\dz u \db v + \db u \dz v)
          \over (1 - u v) (1 + \lambda - u v) } \cr
              & + {1 \over \pi} \ints { 1 - u v \over 1 + \lambda - u v}
                                                          \dz x \db x \cr
              & + {1 \over 2 \pi} \ints { c \over 1 + \lambda - u v}
         ( v \dz u \db x - v \db u \dz x - u \dz v \db x + u \db v \dz x)
           \cr } \eqno(almost)$$
where $\lambda \equiv 2 c^2/k$.
This action can be greatly simplified by making the field redefinition
$$u = e^{\sqrt{2} t/\sqrt{k(1+\lambda)}} \sqrt{\r - (1 + \lambda)}, \>\>
  v = - e^{-\sqrt{2} t/\sqrt{k(1+\lambda)}} \sqrt{\r - (1 + \lambda)} \>\>,
                   \eqno(rtcoord)$$
after which the action becomes
$$\eqalign{ L = {1 \over \pi} \ints &
         {k \dz \r \db \r \over 8 \r^2 (1 - \lambda/\r) (1 - (1 + \lambda)/\r)}
          - \left(1 - {1 + \lambda \over \r}\right) \dz t \db t \cr
          + & \left(1 - {\lambda \over \r} \right) \dz x \db x
 + \sqrt{{\lambda \over 1 + \lambda}} \left(1 - {1 + \lambda \over \r}\right)
                (\dz x \db t - \db x \dz t) \>. \cr } \eqno(nearly)$$
This describes a string propagating in a spacetime with metric
$$ ds^2 = - \left(1 - {1 + \lambda \over \r}\right) dt^2
          + \left(1 - {\lambda \over \r}\right) dx^2
	  +\left(1 - {1 + \lambda \over \r}\right)^{-1}
	  \left(1 - {\lambda \over \r}\right)^{-1}
           {k d\r^2 \over 8 \r^2 }
                                   , \eqno(targmet) $$
and an antisymmetric tensor field
$$B_{tx} = \sqrt{ \lambda \over 1 + \lambda} \left(1 - {1 + \lambda \over \r}
                     \right) \>. \eqno(buv) $$

The exact central charge of this gauged WZW model is $3k/(k-2)$. This
is one larger than the two dimensional black hole since we have added
an extra boson. Eqs.~\(targmet) and~\(buv) yield the lowest order
expressions for the metric and antisymmetric tensor field, but quantum
(sigma model) corrections will introduce corrections.  There is also a
dilaton which arises from quantum effects.  This is most easily
obtained from the requirement that the fields must be an extremum of the low
energy string action
$$S= \int e^\Phi \left[ R + (\nabla\Phi)^2 - {1 \over 12} H^2
                + {8 \over k} \right] \eqno(lowenergy)$$ 
where the cosmological constant $8/k$ arises from the fact that the central
charge is not equal to the spacetime dimension.
It is straightforward to verify that the above metric and antisymmetric
tensor field are indeed an extrema of this action if 
$$ \Phi = \ln \r +a \eqno(dilatona)$$
for an arbitrary constant $a$.

The metric
components~\(targmet) are ill behaved at $\r = 0,\lambda$, and $1 +
\lambda$. We can test whether these are true singularities by looking
at the scalar curvature, which is
$$R = {4 (2 \r + 4 \lambda \r - 7 \lambda - 7 \lambda^2) \over k \r^2 }
                    . \eqno(scalar) $$
Thus $\r=0$ is a curvature singularity. As suggested by eq.~\(scalar),
the difficulties at $\r = \lambda$ and $\r = 1 + \lambda$ can be
completely removed by an appropriate change of coordinates. In fact,
the original $u,v,x$ coordinates in eq.~\(almost) are well behaved at
$uv=0$ which corresponds to $\r = 1 + \lambda$. (The $u,v,x$
coordinates are not well behaved at $uv=1$ ($\r=\lambda$). This is a
direct consequence of the fact that our gauge fixing breaks down
there. Note that unlike the case of the two dimensional black hole,
the spacetime is nonsingular where the gauge fixing breaks down.)  We
will see that $\r = 1+\lambda$ is an event horizon.  The solution is
clearly invariant under translations of both $t$ and $x$, and for
large $\r$ the metric is asymptotically flat.  Thus the solution
represents a straight, static, black string.

We now wish to reexpress the free parameters $\lambda$ and $a$ in
terms of the physical mass per unit length and axion charge per unit
length of the black string. First note that the overall scaling for
the $t$ and $x$ coordinates is fixed by the condition that as $\r$ goes
to infinity, the metric components $g_{tt}$ and $g_{xx}$ approach
unity. It is not possible to similarly fix the overall scaling of the
coordinate $\r$ since the metric asymptotically approaches $k d\r^2/8 \r^2$.
It will be convenient to fix the scaling of $\r$ so that the
dilaton is exactly $\Phi=\ln r + {1 \over 2} \ln {k \over 2}$. In
other words we set
$$\r =r e^{-a} \sqrt{ k \over 2} \eqno(newr) $$ 
in eqs.~\(targmet),~\(buv), and~\(dilatona).
This has the virtue that the metric now depends on two parameters
which we will see are simply related to the physical mass and charge
per unit length. The fact that the metric depends on both the mass and
charge is of course the familiar situation with higher dimensional
black holes and black strings.

The axion charge is computed as follows. In $n$ dimensions, it follows
from the action~\(lowenergy) that the $n-3$ form
$K = {1 \over 2} e^\Phi *H$ is curl free where $*$ denotes the Hodge
dual.  The axion charge per unit length is the integral of this form
over the $n-3$ sphere at large transverse directions from the
string\footnote{$^\dagger$}{This definition differs by a factor of
${1 \over 2}$
from the one used in ref.~[\cite{Horowitz}]}.  Since we are in three
dimensions, $K$ is just a function which must be constant by the field
equations.  The axion charge per unit length is simply the value of
this constant. For the black string solutions we obtain
$$ Q = e^a \sqrt{{2 \lambda (1 + \lambda)\over k}} \>. \eqno(charge)$$

To calculate the mass per unit length of the string we follow the standard
ADM procedure.
For large $r$ the black string solutions approach the
asymptotic solution
$$\eqalign{   ds^2 & = -dt^2 + dx^2 + d\rho^2 \cr
              \Phi & = \rho \sqrt{{8\over k}} \cr} \eqno(asympt) $$
where $\rho = \sqrt{ k \over 8} \ln (r\sqrt{k/2}) $
and $H=0$.  To calculate the mass, one
extremizes the action~\(lowenergy) to obtain the metric field
equation, linearizes this expression about the asymptotic solution~\(asympt),
and
integrates the time-time component of this equation over a constant
time surface. Since the integrand is a total derivative, the result
can be expressed as a surface integral at infinity. The
antisymmetric tensor field appears quadratically in the metric field
equation and vanishes in the background, so it does not explicitly appear
in the formula for the mass. We therefore only need to keep track of
the metric and dilaton. Their contributions to the field equation are
$$  e^\Phi\left[R_{\mu\nu} - \half R g_{\mu\nu} -\d_\mu\d_\nu \Phi +
       g_{\mu\nu}\left( \d^2 \Phi + \half (\d\Phi)^2 - 4/k\right)\right]
                                               . \eqno(metlow)$$
We now linearize this expression. Since $k$ appears in the background 
solution, it cannot be changed by the perturbation\footnote*{It is as 
meaningless to compare the masses of two solutions  with different
$k$, as it is to compare the masses of two Kaluza-Klein solutions with
different compactifications.}.  We have
chosen our radial coordinate so that, in our solutions,
$\Phi$ depends only on $k$. So to calculate their mass, we do not need to
include a perturbation of $\Phi$.
 We need only perturb the metric
$g_{\mu\nu} = \eta_{\mu\nu} + \g_{\mu\nu}$.
Integrating the time-time component of the linearized form of eq.~\(metlow)
over a spacelike surface yields the following formula for the total mass:
$$ {\cal M}_{tot}= \half \oint e^\Phi[ - \p_i \g_{00} +
         \p^\mu\g_{\mu i} -\p_i \g + \g_{i\mu}\p^\mu \Phi] dS^i \>. 
	 \eqno(massform)$$
Using the specific form of $\g_{\mu\nu}$ for our solutions~\(targmet), and
the fact that $x$ measures proper
distance at infinity we obtain the mass per unit length
$$ M = \sqrt{2 \over k} (1+\lambda) e^a \eqno(mass)$$

Combining the above results, we obtain our final
expression for the black string solutions
$$  ds^2 = - \left(1-{M\over r}\right) dt^2
           + \left(1-{Q^2\over M r}\right)dx^2
           + \left(1-{M\over r}\right)^{-1} \left(1-{Q^2\over Mr}\right)^{-1}
             {k \, dr^2 \over 8 r^2} \eqno(final2)$$
with antisymmetric field strength and dilaton
$$ \eqalign{ H_{rtx} & = Q/ r^2 \cr
          \Phi & = \ln r + {1 \over 2} \ln {k \over 2}. \cr} \eqno(handd2) $$
When $Q=0$, $H$ vanishes and
our black string solutions become a simple product of $dx^2$ and the
two dimensional metric
$$ ds^2 = - \left(1-{M\over r}\right) dt^2
          + \left(1-{M\over r}\right)^{-1} {k \, dr^2 \over 8r^2}
                            .  \eqno(wittens) $$
This is exactly Witten's black hole solution~[\cite{Witten}].  It can
be put into his form by the coordinate transformation $r = M \cosh^2 \rho$,
$t = \sqrt{k \over 2} \tau$. The form of the metric~\(wittens) shows clearly
that the region beyond the singularity ($r<0$) has negative mass.

\subhead{\bf 3. Global Structure}

Before discussing the global properties of the metric~\(final2), let us first
review the charged black hole solution in four dimensional general
relativity. This is described by the \RN\  metric
$$  ds^2 =  -\left(1-{2M\over r} + {Q^2 \over r^2}\right) dt^2 +
             \left(1-{2M\over r} + {Q^2 \over r^2}\right)^{-1} dr^2
                     +r^2 d\Omega^2 \eqno(rnmetric)$$
with $|Q| \le M$.  The metric components are ill defined at $r=0$
and $r=r_{\scriptscriptstyle \pm} \equiv M \pm \sqrt{M^2 - Q^2}$. Only
the first is a true curvature singularity.  The surface
$r=r_{\scriptscriptstyle +}$ is the event horizon. The surface
$r=r_{\scriptscriptstyle -}$ is another null surface inside the black
hole called the inner horizon.  It has the following interpretation.
Given initial data for some fields on a $t=$ constant surface, one can
uniquely evolve only up to the inner horizon. After that, the evolution
is affected by the boundary conditions at the singularity.  Since $r=$
constant surfaces are timelike near the singularity, the singularity
is also timelike, in contrast to the familiar Schwarzschild solution.
This means that observers falling into the black hole are not required
to hit the singularity, but can pass by it into another asymptotically
flat region of spacetime. In fact, one can show that freely falling
test particles never hit the singularity: The \RN\ solution
is timelike geodesically complete. Since the metric is
spherically symmetric, its global structure can be described by a two
dimensional figure in which each point represents a sphere.
(Equivalently, one can view the figure as representing the two
dimensional spacetime obtained by holding $\theta$ and $\phi$ fixed.)
It is convenient to rescale the metric so that infinity is at a finite
distance, and have light rays travel along 45 degree lines. The result is
called a Penrose diagram. For the $|Q|<M$ \RN\ solution,
the Penrose diagram  is shown in fig.~1. (The regions VII and VIII
correspond to $r<0$ or equivalently $M<0$ and are usually not included.
They describe asymptotically flat spacetimes with naked singularities.)
The result is an
infinite family of asymptotically flat regions joined by black holes.
For more details see Hawking and Ellis~[\cite{Hawking}]. 
When $|Q|=M$, the horizons coincide, and the
global structure is described in fig.~2. Finally, when $|Q|>M$, the spacetime
contains a naked singularity. The Penrose diagram is equivalent to region
VII of fig.~1.

The inner
horizon is believed to be unstable. It has been shown that generic
time dependent perturbations blow up there~[\cite{Chandra}].  When the
dilaton is included as in string theory, even static spherically
symmetric charged black holes do not have an inner
horizon~[\cite{Gibbons}-\cite{Ishibashi}]. It is thus surprising that the
charged black string does have an inner horizon, as we now discuss.

\subhead{{\bf A.} The black string with $0<|Q|<M$}

Consider the black string solutions~\(final). 
Near $r=M$, the metric is similar to
the region near $r=r_{\scriptscriptstyle +}$ of the \RN\  solution. As in that
case, $r=M$ is an
event horizon. Near $r=Q^2/M$, the metric is similar to the
region near $r=r_{\scriptscriptstyle -}$ and this is indeed an inner horizon.
However there is one important difference. In \RN, the
Killing vector which is timelike at infinity becomes spacelike between
the two horizons and timelike again inside the inner horizon. In the
black string solution, the Killing vector $\p/\p t$ which is a time
translation at infinity becomes spacelike inside the event horizon and
stays spacelike all the way to the singularity. The Killing vector
$\p/\p x$ which translates along the string at infinity becomes
{\it timelike} inside the inner horizon. Equivalently, the time coordinate
in region I of fig.~1 is $t$, the time coordinate in region II is $r$,
and the time coordinate in region V is $x$.\footnote{$^\dagger$}{This
shows that it is not possible to compactify the $x$ direction and view
this as a two dimensional solution.} For this reason it is not
possible to represent all aspects of the causal structure by a two
dimensional diagram.  Nevertheless, most features are faithfully
indicated by fig.~1 with $r_{\scriptscriptstyle +}= M$ and
$r_{\scriptscriptstyle -}= Q^2/M$.  
Each point now represents a line in spacetime.
However it is the line in the $x$ direction for $r>Q^2/M$ and
in the $t$ direction for $r<Q^2/M$. Unlike \RN,  the black string solution also
contains the regions labeled VII and VIII in fig.~1 which correspond
to naked singularities.  This is because these regions correspond to
$r<0 $ or
$uv<-(1+\lambda)$ which is certainly part of the original gauged WZW model.

We now consider geodesics in the black string solutions. This is
particularly simple due to the two conserved quantities associated
with the two translational symmetries. Let $\xi^\mu$ be tangent to
an affinely parametrized geodesic, and let $E = -\xi \cdot \p/\p t$,
$P = \xi \cdot \p/\p x$ denote the conserved quantities. Then
geodesics satisfy
$$  { k \dot r^2 \over 8r^2 } = E^2-P^2
                 + {1\over r} \left( P^2 M - {E^2 Q^2 \over M}\right)
      +\alpha (1- M/r)(1-Q^2/Mr)  \eqno(geos) $$
where the dot denotes derivative with respect to an affine parameter
and $\alpha$ is zero for null geodesics and $-1$ for timelike
geodesics.  In either case, if the right hand side is positive for
large $r$, it stays positive for all $r> Q^2/M$ . This shows that
timelike and null geodesics which begin at large $r$ cross both the
event horizon at $r=M$ and the inner horizon at $r=Q^2/M$. For
timelike geodesics, the term $-Q^2/r^2$ eventually dominates causing
$r$ to achieve a minimum value.  Thus timelike geodesics never reach
the singularity: The spacetime is timelike geodesically complete.
Eq.~\(geos) has a simple solution for null geodesics. If $P^2 < E^2$, then
$$ r = { (e^{\lambda} - P^2 M^2 + E^2 Q^2)^2 \over
          4 e^{\lambda} M (E^2 - P^2) } \,, \eqno(rnull) $$
where $\lambda$ is an affine parameter. If $P^2 = E^2$, then $r = \lambda^2$.
Null geodesics can reach the singularity, but only if $P^2 M^2 - E^2 Q^2 >0$.
Otherwise, the null geodesics reach a minimum value of
$$r_{min} = {E^2 Q^2 - P^2 M^2 \over M (E^2 - P^2) }\,. \eqno(rmin) $$
This is qualitatively the same behavior as geodesics in
the \RN\  solution.

There appears to be a problem with geodesics having $P=0$. These
geodesics are orthogonal to $\p/\p x$ everywhere, but inside the inner
horizon this vector becomes timelike and cannot be orthogonal to any
timelike or null geodesic. As we can see from eqs.~\(geos)
and~\(rmin), the resolution is that geodesics with $P=0$ cross the
point labeled p in fig.~1 (recall this is really a line in spacetime).
At p the Killing vector $\p/\p x$ is not only null, but
actually vanishes.

We have seen that the three dimensional black string is qualitatively
very similar to the \RN\  solution. This analogy appears
to extend to Hawking evaporation. We can define a Hawking temperature
for the black string by analytically continuing $t =i\tau$ in
eq.~\(final).  The horizon $r=M$ is a regular point only if we
identify $\tau$ with period $\pi M \sqrt{2 k/ (M^2-Q^2)}$ which
corresponds to a temperature
$$ T  = {1\over \pi M} \sqrt{{M^2 - Q^2 \over  2 k}} \>. \eqno(temp) $$
Therefore, the temperature vanishes as $Q\a M$. 
(This is also true of \RN.)
Thus, if the charge cannot be radiated away, the black strings
would settle down to
$|Q|=M$.

\subhead{{\bf B.} The extremal limit: $|Q|=M$}

What does the extremal configuration look like?  If one sets
$|Q|=M$ in eq.~\(final) one obtains
$$ ds^2 = (1-M/r)(-dt^2 + dx^2)+ (1-M/r)^{-2} k\, dr^2/8r^2 \>. \eqno(limit) $$
Notice that this extremal metric is not only static and
translationally invariant, it is also boost invariant along the
string.  (Higher
dimensional extended black holes are also boost invariant in the extremal
limit~[\cite{Horowitz}].)  At first sight, the global structure of the
metric~\(limit)
appears to be analogous to the extreme \RN\  metric with
a single horizon at $r=M$ and a singularity at $r=0$.  However this is
misleading. The proper continuation across the horizon is {\it not} to let
$r$ become less than $M$. This is most easily seen by returning to the
geodesic equation~\(geos). When $|Q| = M$ this equation becomes
$$  { k \dot r^2 \over 8r^2 } = (E^2-P^2)(1-M/r) 
      +\alpha (1- M/r)^2 \>.\eqno(geoslimit) $$
Near the horizon, the second term on the right 
is negligible and the first changes sign. Thus no timelike or null geodesics
reach $r<M$. To find the correct extension across the horizon we must 
introduce a new radial coordinate. Starting in the region outside the
horizon, set
$$  \t r^2 = r-M \eqno(rlimit)$$
Then the metric becomes
$$ds^2 ={\t r^2 \over \t r^2 +M} (-dt^2 + dx^2) + {k \, d\t r^2 \over 2 \t r^2}
           \>. \eqno(metlimit)$$
One can easily verify that geodesics now cross the horizon $\t r=0$ from
positive to negative values of $\t r$ so the metric~\(metlimit) and
not~\(limit) describes the correct extension across the horizon. But
the region $\t r <0$ is identical to the region $\t r >0$, and the
metric~\(metlimit) is nonsingular! Nevertheless, $\t r=0$ is still an
event horizon. Thus one has the unusual situation of a spacetime with
an event horizon but no singularity\footnote*{The authors of
ref.~[\cite{Ishibashi}] have found  a two dimensional solution
with similar properties  by modifying the
construction described in their paper.}.
The global structure is described in fig.~3. Observers in this spacetime
who cross the event horizon  are not able to return, but (fortunately for
them) find themselves in another asymptotically flat region of spacetime 
which is identical to the one they started in. 

What happens to the region V in fig.~1, near the singularity, 
as $|Q|$ approaches $M$?
Setting $\t r^2 = M-r$, the metric in this region becomes
$$ds^2 ={\t r^2 \over M-\t r^2 } (dt^2 - dx^2) + { k \, d\t r^2 \over 2 \t r^2}
                     \eqno(inside) $$
which is singular at $\t r= \pm \sqrt M$ and has a horizon
at $\t r = 0$. The causal structure looks like fig.~4.

There is in fact another way to take the extremal limit. To motivate it, let
us return to the conformal field theory construction. One starts with
$\slr \times \R$ and gauges a one dimensional subgroup of $\slr$
and a translation in $\R$. The free parameter $c$
determines the relative weights of the two gaugings. One limiting case
($c=0$), corresponds to not gauging $\R$ at all. This yields the
uncharged black string. The other limit ($c \a\infty$) corresponds to
not gauging the $\slr$. The result is just the group invariant metric
on $\slr$ which has constant negative curvature and hence is three
dimensional anti-de Sitter space. But this limit also corresponds to
$|Q| \a M$. Thus the ungauged $\slr$ WZW model can be viewed as
describing the extremal black string. Notice that this solution has
constant dilaton, and is not asymptotically flat. It can be obtained
from the metric~\(final) by rescaling
$y = {8 (r-M)\over k (M^2 - Q^2)^{1/2}}, \hat t = (1-Q^2/M^2)^{1/4} t,
\hat x = (1 - Q^2/M^2)^{1/4} x$ and then taking the
limit $|Q| \a M$\footnote{$^\dagger$}{This extremal limit is analogous
to the one
used in ref.~[\cite{Giddings}] for black fivebranes.}. The resulting metric is
$$ ds^2 = {k \over 8} \left(-y\, d\hat t^2 + y \,d\hat x^2 +
                 {dy^2 \over y^2 }\right)    \>. \eqno(antides)  $$
To see that this is indeed anti-de Sitter spacetime (albeit in unusual
coordinates) one can calculate the curvature and find
$R_{\mu\nu} = - {4 \over k} g_{\mu\nu}$. The limiting antisymmetric tensor
is simply $H =\epsilon \sqrt{8/k}$  where $\epsilon$ is the volume form.

In ten dimensions, the extremal limit of the black string
solutions~[\cite{Horowitz}] is of particular interest. It agrees
precisely with the solution found by Dabholkar {\it et al.}~[\cite{Dabholkar}]
describing the fields outside of a fundamental
macroscopic string.  Dabholkar {\it et al.} also found the fields outside
of a fundamental macroscopic string in any dimension. In three
dimensions, their solution is\footnote*{They chose to work with the
Einstein metric and different scaling for the dilaton. We have
reexpressed their solution in terms of the string metric and dilaton
used throughout this paper.}
$$\eqalign{ ds^2 = & {1\over \t y} (-d\t t^2 +d \t x)^2 + d \t y^2 \cr
       \Phi = & \ln \t y \cr
      H = & {\epsilon \over \t y} \> . \cr} \eqno(dab) $$
This solution, like the ten dimensional analog, has a singularity at $\t y=0$
and no horizon. It does not resemble the extremal limit of our black string.
However, in order to compare the two solutions, we must take into account
the fact that we have included a correction to the central charge proportional
to $1/k$ which was not included in ref.~[\cite{Dabholkar}]. If one wants to
view the three dimensional black string, by itself,  as a solution
to critical string
theory, it is necessary to include this modification to the central charge. 
However if one wants to add an internal conformal field theory,
then the central charge need not be modified.
To compare the two solutions,
we must take the limit as $k \a \infty$. Since large $k$ means
that the metric is rescaled by a large factor, the limit $k \a \infty$ 
is usually thought to yield a flat metric.  However, if one starts with
a singular solution, one can take $k\a \infty$ staying close to the singularity
and obtain a nontrivial limit. More precisely, set
$\t y = (k/8)^{1/2} r/M$, $ \t t = (k/8)^{1/4}x$, $ \t x = (k/8)^{1/4} t$
in eq. \(limit). Then in the limit $k\a \infty$, the solution agrees exactly
with eq.~\(dab). 

\subhead{{\bf C.} The solutions with $|Q| >M$}

We now consider the solution when $|Q| >M$. The
metric~\(final) appears to change signature at $r=Q^2/M$. But this is
just another indication that an incorrect extension is being used. The
correct extension can again be found by considering the motion of geodesics.
It corresponds to setting $\t r^2 = r-Q^2/M$. In terms of $t, x, \t r$ 
the metric becomes
$$ds^2 = - {Q^2-M^2+M \t r^2 \over Q^2 + M \t r^2} dt^2
         + {M \t r^2 \over Q^2 + M \t r^2} dx^2
         + {M k \over 2( Q^2 - M^2 + M \t r^2) } d\t r^2 \> . \eqno(qbig) $$
This metric, with $0\le \t r < \infty$, is globally static. It 
has no horizons and no curvature singularity. It
does have a conical singularity at $\t r=0$ which can be removed by
identifying $x$ with period $ \pi Q \sqrt{2 k/(Q^2 - M^2)}$.
The resulting spacetime is completely nonsingular. Notice that
the identification changes the structure of the spacetime at infinity
from $\R^3$ to $\R^2 \times
S^1$. A spacelike surface  $t=$ constant  now resembles an infinite cigar.
This is reminiscent of the form of a Euclidean two dimensional black hole.
In fact if we take the limit $M\a 0$, $Q\a 0 $ keeping $Q^2/M = m$ fixed,
one finds that~\(qbig) reduces to exactly the product of $-dt^2$ and the two
dimensional Euclidean black hole discussed in ref.~[\cite{Witten}].

This insight helps us to resolve another aspect of these solutions. The 
conformal field theory construction described in the previous section
only yields the
solutions~\(final) with $|Q| <M$. However, the fields~\(qbig) with $|Q| >M$
also solve the low energy string equations and its natural to ask what
is the exact conformal field theory that they correspond to. The answer
is a slight modification of the construction in sec.~2. One again starts
with $\slr \times \R$ but now puts a timelike metric on $\R$. One then gauges
a translation of $\R$ together with the subgroup of $\slr$ generated by
$\pmatrix{0&1\cr-1&0}$.
The result is exactly the solutions~\(qbig) with $|Q|>M$. In the limit that
$\R$ is not gauged, one obtains the two dimensional Euclidean black hole
cross $-dt^2$.

Investigations of the two dimensional Euclidean black hole have shown
that this nonsingular space is dual to the Euclidean negative mass
solution which has a curvature
singularity~[\cite{Giveon},\cite{Kiritsis},\cite
{Tseytlin},\cite{Dijkgraaf},\cite{Martinec},\cite{Bars}].
In other words, these two different geometries are equivalent as
conformal field theories. Although this sounds intriguing, the
physical interpretation of this result remains unclear. The reason is
that in the Euclidean context, the duality involves string winding
modes in Euclidean time. In the Lorentzian context, both the original
conformal field theory and its dual contain all six regions of the
black hole, so the spacetime metric does not change at
all!\footnote*{We thank E.~Verlinde and H.~Verlinde for a discussion
on this point.} In three dimensions there is an analog of this duality
with a clear physical interpretation and a rather striking conclusion.
By the usual two dimensional arguments, the spacetime
$$  ds^2 = -dt^2 
          + \left(1-{M\over r}\right)^{-1} {k \, dr^2 \over 8r^2}
 +\left(1-{M\over r}\right) d\theta^2
 \eqno(euplus)$$
(where the spatial metric -- with $r\ge M$ -- is the positive mass
Euclidean black hole)
is completely equivalent as a conformal field theory to the spacetime
$$  ds^2 = -dt^2 
          + \left(1+{M\over r}\right)^{-1} {k \, dr^2 \over 8r^2}
 +\left(1+{M\over r}\right) d\theta^2
 \eqno(euminus)$$
(where the spatial metric is now the negative mass Euclidean
solution).  The duality is now the familiar one involving winding
modes in a spacelike direction.  There is no need for further analytic
continuation.  This shows that the curvature singularity in
eq.~\(euminus) is not seen by strings. String propagation in this
spacetime is completely equivalent to string propagation in the
nonsingular geometry~\(euplus). This duality can be extended to the
general solution with $|Q| > M$, and will be discussed in detail
elsewhere~[\cite{dualstuff}].

\subhead{\bf 4. Conclusions}

We have been describing the black string 
solutions in terms of the metric which appears
in the sigma model. This is the metric that the strings couple directly to
and is the most natural one to use in string theory. However to compare
with results in general relativity, it is sometimes useful  to rescale this
string metric by a power of the dilaton to obtain a metric with the standard
Einstein action. Since conformal transformations do not change the causal
structure, both the original string metric and the new Einstein metric will
have the same horizons. But since the dilaton is growing linearly at infinity
one might think that the Einstein metric will not be asymptotically flat. This
is incorrect. In three dimensions, the Einstein metric $\tilde g_{\mu\nu}$
is related to the string metric by $\tilde g_{\mu\nu}= e^{2\Phi}g_{\mu\nu}$.
Thus asymptotically, the Einstein metric approaches
$$ \tilde ds^2 = r^2(-dt^2 +dx^2) + {k \over 8} dr^2 \> . \eqno(einstein)$$
One can check that the curvature of this metric vanishes at large $r$ like
$r^{-2}$. In $n$ dimensions, the usual definition of asymptotic flatness
requires that the curvature fall off like $r^{-(n-1)}$. So it is  
reasonable to view eq.~\(einstein) as an asymptotically flat three dimensional 
spacetime\footnote*{In higher dimensions, if one starts with a 
flat string metric and a
linear dilaton, and rescales to the Einstein metric, one again finds that
the curvature falls off like $r^{-2}$. Thus higher dimensional analogs
could not be considered asymptotically flat in the usual sense.}.
Thus the Einstein metric also describes a black string. But it is not static.
The timelike symmetry  at large $r$ does not approach a time translation
but rather a boost. One can check that
the Einstein metric is still singular at $r=0$ (for $|Q|<M$).

The fact that the Einstein metric describes a {\it nonstatic} 
string resolves an
apparent contradiction with an earlier result on the existence of
black strings in low dimensions. It was shown in ref.~[\cite{Horowitz}]
that if the strong energy condition is satisfied, there cannot exist
static, cylindrically symmetric, black strings in less than five dimensions
which are locally asymptotically flat in the transverse directions. This
does not rule out three (or four) dimensional 
solutions in which the string metric
describes static black strings, since the  low energy field theory
does not satisfy the energy condition. However, when expressed in terms of
the Einstein metric, the low energy field theory {\it does} satisfy the
energy condition. Since the causal structure is unchanged, and the asymptotic
boundary conditions on the metric are unchanged, the only way these solutions
can exist is if they are not static. It would appear that they cannot even
``settle down" to a static configuration unless they evaporate completely
by Hawking radiation. (Recall that the extremal case with zero temperature
still has a horizon and hence cannot be static.)  

To summarize, we have found that a simple gauged WZW model yields
charged black string solutions in three dimensions. When $0<|Q|<M$,
the solutions are qualitatively similar to the \RN\ solution. They
have an event horizon, an inner horizon and a timelike
singularity. When $|Q| = M$ the spacetime has an event horizon but no
singularity. (Another way to take the extremal limit, which does not
preserve the boundary conditions, yields anti-de Sitter spacetime.)
When $|Q|>M$, both the horizon and the curvature singularity
disappear. To avoid a conical singularity one must compactify one of
the directions. A limiting case yields just the product of time and
the two dimensional Euclidean black hole.

Since our black string solutions are three dimensional and have a
linear dilaton at infinity, they presumably are not of direct physical
interest.  Their importance is twofold. First, they illustrate that a
wide range of causal structures (including some having no analog in
general relativity) can occur in string theory. Indeed, we find it
surprising that a simple conformal field theory construction can
result in such nontrivial spacetime structure. This encourages the
hope that an exact conformal field theory describing higher
dimensional black holes and black strings will soon be found.  Second,
like the two dimensional black hole, they provide an important test of
whether gravitational collapse will lead to singularities in string
theory. It has been shown that string theory does have exact solutions which
are singular~[\cite{Steif}].
However the known singular solutions do not have event
horizons and hence do not describe gravitational collapse. 

It is still
not clear whether the two dimensional black hole is singular in string
theory.  (Recall that although one has an exact description of the
conformal field theory in terms of a gauged WZW model, the spacetime
metric~\(wittens) is only the lowest order approximation to the
geometry.  Higher order corrections can be large near the
singularity.) It has been argued~[\cite{Witten}] that even though the 
conformal field theory may be regular at $r=0$, it does not make sense to
consider signals propagating from $r>0$ to $r<0$ for two reasons. One is that,
in the two dimensional black hole,
the surface $r=0$ is spacelike on one side and
timelike on the other. Thus if signals can propagate across, there
would appear to be a violation of causality. The other is that $r=0$ appears
to be unstable in that generic perturbations blow up there.
For our three dimensional black strings (with
$0<|Q|<M$) the surface  $r=0$ is timelike on
both sides,  so no causality problems should arise. On the other hand,
general arguments suggest that the inner horizon  $r=Q^2/M$ is now 
unstable. So once again,  it appears to be impossible to propagate signals
from large positive $r$ to large negative $r$.

\subhead{Acknowledgements}
We wish to thank S.~Giddings, N.~Ishibashi, M.~Li, A.~Steif,
A.~Strominger, and especially D.~Garfinkle for helpful discussions.
J.H.H.~would like to thank the Aspen Center for Physics, where this
work was begun. This work was supported in part by NSF grant
PHY-9008502.

\references

\baselineskip=16pt

\refis{Witten} E.~Witten, ``On String Theory and Black Holes,''
\journal Phys. Rev., D44, 314, 1991.

\refis{Dijkgraaf} R.~Dijkgraaf, E.~Verlinde and H.~Verlinde,
``String Propagation in a Black Hole Geometry,'' Princeton
preprint, PUPT-1252, May 1991.


\refis{Tseytlin} A.~Tseytlin, ``Space-Time Duality, Dilaton, and
String Cosmology,'' to appear in {\it Proceedings of
the First International Sakharov Conference on Physics}, May 1991.

\refis{Martinec} E.~Martinec and S.~Shatasvili, ``Black Hole Physics
and Liouville Theory,'' Enrico Fermi preprint, EFI-91-22, May 1991.

\refis{Kiritsis} E.~Kiritsis, ``Duality in Gauged WZW Models,''
Berkeley preprint, LBL-30747, May 1991.

\refis{Bars} I.~Bars, ``Curved Space-Time Strings and Black Holes,``
USC preprint, USC-91-HEP-B4, June 1991;
``String Propagation on Black Holes,'' USC preprint, USC-91-HEP-B3, May 1991.

\refis{Giddings} S.~Giddings and A.~Strominger, ``Exact Black Fivebranes
in Critical Superstring Theory,'' UCSB preprint, UCSBTH-91-35, July 1991.

\refis{Giveon} A.~Giveon, ``Target Space Duality and Stringy Black Holes,''
Berkeley preprint, LBL-30671, April 1991.

\refis{Ishibashi} N.~Ishibashi, M.~Li and A.~Steif,
``Two Dimensional Charged Black Holes in String Theory,''
UCSB preprint, UCSBTH-91-23, July 1991.

\refis{Gawedzki} K.~Gawedzki and A.~Kupiainen,
\journal Phys.~Lett., B215, 119, 1988; \np B320, 625, 1989.

\refis{Horowitz} G.~Horowitz and A.~Strominger, ``Black Strings
and $p$-Branes,''
\journal Nucl.~Phys., B360, 197, 1991.


\refis{Hawking} S.~Hawking and G.~Ellis, {\it The Large Scale Structure
of Space-Time} (Cambridge Univ.~Press, Cambridge) 1973.

\refis{Gibbons} G.~Gibbons,
\journal Nucl.~Phys., B207, 337, 1982.

\refis{Chandra} S.~Chandrasekhar and J.~Hartle,
\journal Proc. Roy. Soc. Lond., A384, 301, 1982.

\refis{Garfinkle} D.~Garfinkle, G.~Horowitz and A.~Strominger, ``Charged
Black Holes in String Theory,''
\journal Phys.~Rev., D43, 3140, 1991.

\refis{Steif} G.~Horowitz and A.~Steif,
\journal Phys.~Rev.~Lett., 64, 260, 1990;
\journal Phys.~Rev., D42, 1950, 1990;
\journal Phys.~Lett., B258, 91, 1991.

\refis{Nemeschansky} I.~Bars and D.~Nemeschansky,
``String Propagation in Backgrounds with Curved Space-time,''
\journal Nucl.~Phys., B348, 89, 1991.

\refis{dualstuff} J. Horne, G. Horowitz, and A. Steif, to appear.

\refis{Maeda} G.~Gibbons and K.~Maeda,
\journal Nucl.~Phys., B298, 741, 1988.

\refis{Ivanov} B.~Ivanov, ``Black Holes and the Heterotic String,''
ICTP preprint, IC/89/3, January 1989.

\refis{Dabholkar} A.~Dabholkar, G.~Gibbons, J.~Harvey and F.~Ruiz,
``Superstrings and Solitons,''
\journal Nucl. Phys., B340, 33, 1990.

\endreferences

\figurecaptions

\baselineskip=16pt

\item{Figure 1:} The global structure for both the \RN\ solution
and for the black string when $0<|Q|<M$. The jagged lines represent
singularities, and $r=r_{\scriptscriptstyle \pm}$ represent horizons.
For the \RN\ solution, 
$r=r_{\scriptscriptstyle \pm} \equiv M \pm \sqrt{M^2 - Q^2}$ and
each point in the figure represents a sphere in spacetime. 
For the black string, $r_{\+} = M$ and $r_{\-} = Q^2/M$.
Each point in the figure represents a line in spacetime. 
Geodesics with $P=0$ cross the point $p$
in the diagram. (The regions VII and VIII correspond to $r<0$ and
are usually not considered part of the \RN\ solution.) 

\item{Figure 2:} The global structure for the extreme \RN\ solution,
$|Q| = M$. The two horizons in fig.~1 have become
a single horizon.

\item{Figure 3:} The global structure for the extreme black string solution,
$|Q| = M$, keeping the region $r \rightarrow \infty$.
The spacetime has  an event horizon, but no singularities.

\item{Figure 4:} The global structure for the extreme black string solution,
$|Q| = M$, keeping the region near the singularity.

\endfigurecaptions

\endit\end
\end